# CNN Based Detection of Cardiovascular Diseases from ECG Images

1st Irem Sayin
*Yildiz Technical University*
*Mechatronics Engineering*
Istanbul,Turkey

2nd Rana Gursoy
*Yildiz Technical University*
*Mechatronics Engineering*
Istanbul,Turkey

3rd Buse Cicek
*Yildiz Technical University*
*Mathematics and Science Education*
Istanbul,Turkey

4th Yunus Emre Mert
*Yildiz Technical University*
*Mechatronics Engineering*
Istanbul,Turkey

5th Fatih Ozturk
*Yildiz Technical University*
*Mechatronics Engineering*
Istanbul,Turkey

6th Taha Emre Pamukcu
*Yildiz Technical University*
*Electronics and Communication Engineering*
Istanbul,Turkey

7th Ceylin Deniz Sevimli
*Saint Benoit Lycee*
Istanbul,Turkey

8th Huseyin Uvet
*Yildiz Technical University*
*Mechatronics Engineering*
Istanbul,Turkey

*Abstract-* This study develops a Convolutional Neural Network (CNN) model for detecting myocardial infarction (MI) from Electrocardiogram (ECG) images. The model, built using the InceptionV3 architecture and optimized through transfer learning, was trained using ECG data obtained from the Ch. Pervaiz Elahi Institute of Cardiology in Pakistan. The dataset includes ECG images representing four different cardiac conditions: myocardial infarction, abnormal heartbeat, history of myocardial infarction, and normal heart activity. The developed model successfully detects MI and other cardiovascular conditions with an accuracy of 93.27%. This study demonstrates that deep learning-based models can provide significant support to clinicians in the early detection and prevention of heart attacks.

## I. Introduction

Cardiovascular diseases are one of the leading causes of death and morbidity worldwide [1]. According to the World Health Organization, an estimated 17.9 million people died from cardiovascular disease in 2019 [1]. Among the clinical manifestations of cardiovascular diseases, myocardial infarction holds a significant place. Myocardial infarction typically presents as an intense pain or pressure in the center of the chest, which may radiate to the arms, left shoulder, jaw, or back. Such symptoms are considered a serious indication of underlying cardiovascular pathology [1].

In recent years, the use of machine learning and deep learning techniques in the diagnosis of heart diseases has become increasingly prevalent [2-4]. While traditional diagnostic methods include techniques such as electrocardiograms (ECG) and magnetic resonance imaging (MRI), these new approaches offer significant advantages in the diagnostic process [5]. Deep learning methods, in particular, provide more novel and effective solutions that require automatic feature extraction from ECG data.

Convolutional Neural Networks (CNN), a specialized type of deep learning algorithm, are utilized to better and more rapidly predict heart diseases and abnormalities based on ECG data [6]. CNNs and other deep learning techniques hold a significant place in the field of medical imaging, particularly excelling in large datasets. These algorithms are widely used for identifying critical patterns and structures in image data and for analyzing medical images [7, 8]. They offer high accuracy rates in the analysis of medical signals [9].

The use of artificial intelligence and deep learning algorithms in the analysis of cardiovascular diseases has the potential to enhance the accuracy and effectiveness of medical decision support systems. The performance of artificial neural networks in detecting arrhythmias using ECG signals has been compared to that of cardiologists, with similar or even higher accuracy rates observed in some cases [10]. For example, meaningful features have been automatically learned and classified from 12-lead ECG images, both in color and grayscale, without the need for manual feature extraction [11].

In their 2018 study, Liu et al. developed a CNN model for detecting myocardial infarction using multi-lead ECG data, achieving a sensitivity of 95.40%, a specificity of 97.37%, and an accuracy rate of 96.00% [12].

In their 2017 study, Acharya et al. demonstrated that a CNN-based deep learning model achieved accuracy rates of up to 94.90% in classifying rhythm disorders such as atrial fibrillation (A-fib), atrial flutter (A-fl), and ventricular fibrillation (V-fib) using ECG signals segmented into two and five-second intervals [13]. This approach suggests that deep learning could be an effective method for supporting clinical diagnosis processes in the analysis of ECG signals.

Deep learning models can be effective not only in detecting myocardial infarction (MI) but also in diagnosing other cardiovascular diseases [14]. These methods contribute to saving both time and resources, thereby reducing the workload of healthcare professionals.

These findings indicate that the use of deep learning methods in the analysis of myocardial infarction (MI) using ECG signals could overcome the limitations of existing automated ECG analysis systems, leading to more reliable and faster results in clinical applications.

For these reasons, this study utilized an open-source dataset containing ECG images to develop a deep learning model capable of detecting myocardial infarction, normal ECGs, and abnormal ECGs, demonstrating the effectiveness of deep learning methods in improving ECG analysis for clinical applications.

## II. METHODS

### A. Dataset

The dataset employed in this study comprises a total of 929 ECG images, categorized into four distinct classes, each representing a specific cardiac condition: Myocardial Infarction, Abnormal Heartbeat, History of Myocardial Infarction, and Normal Heart Activity. These data were collected from 929 patients, with each patient contributing recordings from 12 ECG leads. Specifically, the dataset includes 240 images from patients diagnosed with Myocardial Infarction, 233 images from patients exhibiting an Abnormal Heartbeat, 172 images from patients with a History of Myocardial Infarction, and 284 images from individuals with Normal Heart Activity. The balanced distribution of these classes ensures a comprehensive representation of the conditions, which is crucial for developing a model capable of generalizing well across various cardiac scenarios. This dataset, created under the auspices of Ch. Pervaiz Elahi Institute of Cardiology Multan, Pakistan, stands as a valuable resource for advancing cardiovascular research [15].

### B. Data Preprocessing

A series of image processing steps were undertaken to enhance data accuracy and clarify signal details. Initially, the images were cut to remove any areas outside the signal region of the ECG paper, focusing solely on the relevant signal data. The images were then loaded in grayscale format to facilitate further processing.

Subsequently, Otsu's thresholding method was applied to convert the grayscale images into a binary format, emphasizing the signal lines by setting a global threshold that separates the ECG signal (white lines) from the background. To refine the images, morphological opening was employed using a disk-shaped structuring element to remove small black dots and noise.

Next, morphological dilation was applied to thicken the ECG signal lines, followed by erosion to enhance the clarity of the signal by eliminating any remaining noise. The processed images were then inverted, resulting in white signal lines on a black background. Finally, the processed images were saved and used the primary dataset for all subsequent analyses and training processes. Following the completion of all preprocessing steps, a representative example of a dataset image before and after processing can be seen in Figure 1.

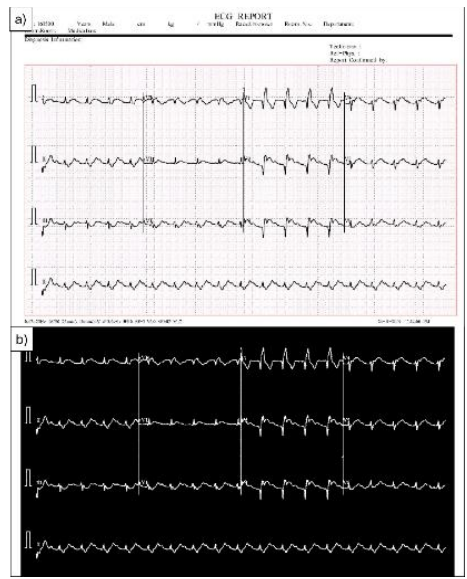

Fig. 1. a) Before preprocessing and b) after preprocessing.

### C. CNN Model

. CNNs require less preprocessing compared to other methods in the field of image processing, making them one of the most effective learning algorithms [16]. Thanks to convolutional and pooling layers, CNNs excel in tasks such as image classification and recognition [17]. Convolutional layers enable better understanding of images by learning both low- and high-level features. Pooling layers, on the other hand, reduce the size of the data, thereby decreasing computational load, accelerating training time, and mitigating the risk of overfitting [16].

Once the feature maps are obtained, a pooling (subsampling) layer is added alongside the convolution layer in CNNs. The purpose of the pooling layer is to reduce the spatial dimensions of the convolved features. This dimensionality reduction decreases the computational power required to process the data. Additionally, this process facilitates the extraction of essential features that are spatially and rotationally invariant, preserving the practical training of the model. Pooling not only shortens the training time but also reduces the risk of overfitting

### D. Inception V3 Model

The Inception V3 model was developed to provide high performance and efficiency in image classification. As illustrated in Figure 2, the architecture of the Inception V3 model involves breaking down large convolutions (such as 7x7) into smaller and asymmetrical convolutions (such as 3x3 and 1x7). This technique significantly reduces computational costs while enhancing the model's learning capacity. [18].

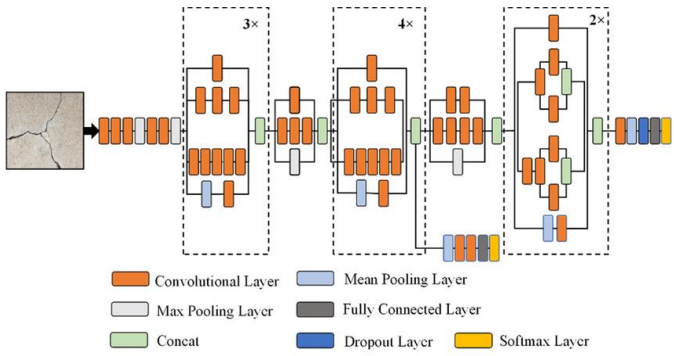

Fig. 2. Inception V3 Model Architecture [19]

Additionally, techniques used to reduce grid sizes facilitate the network's ability to reach deeper layers. The use of auxiliary classifiers helps stabilize the training process by preventing gradient loss. These innovations make Inception V3 a highly efficient model that is widely used in modern image classification projects [20].

*E. Optimizing InceptionV3 with Transfer Learning*

In this study, an InceptionV3-based model was retrained for a specific classification task using transfer learning. The pretrained weights on the ImageNet dataset [21] were retained, with all but the final three layers frozen to preserve the model's strong feature extraction capabilities. The input dimensions were set to 299x299 pixels to match the InceptionV3 architecture. To adapt the model for the classification task, a GlobalAveragePooling2D layer was added to convert the spatial dimensions into dense feature vectors, followed by a Dense layer for final classification.

The final classification was performed using a Dense layer with a Softmax activation function, enabling the model to output probability distributions across classes. After equalizing the class sizes to match the smallest class (172 samples with a history of myocardial infarction), the dataset was then split into 70% training, 15% validation, and 15% testing sets. The model was trained for 200 epochs using hyperparameters optimized through Grid Search, which identified a batch size of 32, learning rate of 0.01, unfrozen layers set to -10, and a dropout rate of 0.5 as the best combination.

Several callback mechanisms were employed during training to enhance model performance and mitigate overfitting. EarlyStopping was utilized to halt training if validation loss did not improve for 10 consecutive epochs, and ModelCheckpoint saved the best models based on validation accuracy and loss. These measures contributed to developing a robust and generalizable classifier.

## III. RESULTS

The InceptionV3-based model demonstrated strong performance on the test dataset, achieving an overall accuracy of 93.27%. Trained over 25 epochs, the model effectively generalized to unseen data, maintaining high precision, recall, and F1 scores, each at 93.27% when averaged using the micro method. This indicates balanced performance across all classes.

The macro average metrics, with a precision of 93.61%, recall of 93.27%, and F1 score of 93.19%, further highlight the model's balanced classification capabilities. The final test loss was 0.2413, further supporting the model's strong generalization ability. Despite class imbalances, the model-maintained accuracy across all categories, demonstrating its robustness after extensive training.

The analysis of the confusion matrix in Figure 3 reveals specific insights into the model's classification behavior. Notably, the "History of Myocardial Infarction" and "Normal" classes were classified with 100% accuracy, displaying the model's exceptional ability to correctly identify these critical categories. On the other hand, the "Myocardial Infarction" and "Abnormal Heartbeat" classes, while still classified with high accuracy, showed some level of misclassification, with true positive rates of 88.46% and 84.62%, respectively. These results suggest that while the model excels in certain areas, there is still room for improvement in distinguishing between these particular conditions.

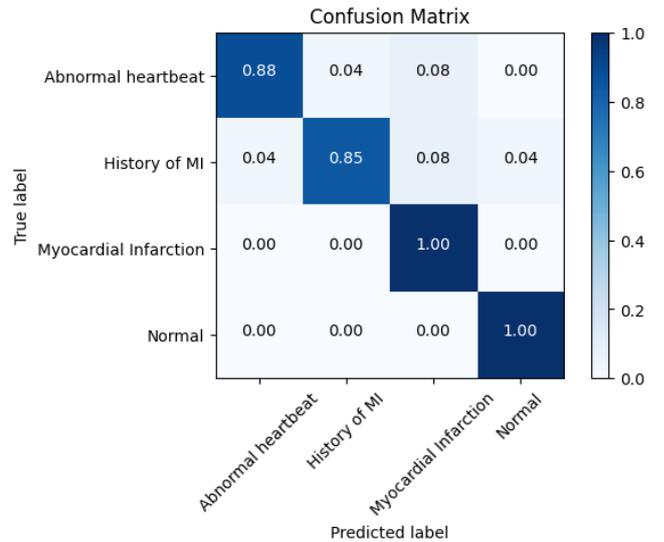

Fig. 3. Confusion Matrix of InceptionV3-Based Model Performance

The heatmaps illustrating the regions of focus during the model's classification process provide valuable insights into the decision-making mechanism of the model, revealing which features were deemed significant. These visualizations demonstrate the model's capability to accurately identify clinically relevant regions, underscoring its potential as a reliable diagnostic tool.

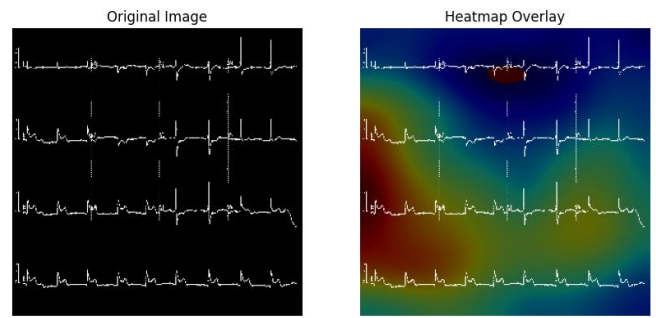

Fig. 4. Heatmap Overlay on Original ECG Image

Figure 4 presents a visual representation of the model's attention during the classification of an ECG image. On the left, the original ECG image is shown, depicting the raw data input to the model. On the right, the heatmap overlay highlights the regions of the ECG that the model focused on most intensely during the classification process. The areas marked in red and yellow in the heatmap indicate the segments of the ECG that had the highest influence on the model's decision, suggesting these areas contained critical features relevant to the classification task. This visualization is crucial for understanding how the model interprets the ECG signals and identifies clinically significant patterns that contribute to accurate diagnosis.

## IV. DISCUSSION

The InceptionV3-based model showed strong performance in classifying cardiac conditions, particularly excelling in identifying normal heart activity and historical myocardial infarction cases. Its high overall accuracy and consistent performance across different classes suggest it could be a valuable diagnostic tool in clinical settings.

However, the model faced challenges in distinguishing between myocardial infarction and abnormal heartbeat conditions, indicating the need for further refinement. Improving the model's sensitivity to these subtle differences might involve more advanced feature extraction techniques or incorporating additional physiological data.

The heatmap visualizations clarify the model's decision-making process, highlighting the most influential ECG signal areas. This transparency could facilitate the model's integration into clinical practice, helping physicians interpret complex ECG patterns more effectively.

Future work should focus on expanding the dataset to include more cardiac conditions to enhance the model's generalizability. Additionally, developing the model into a decision support system could improve the accuracy and efficiency of diagnosing cardiovascular diseases, leading to better patient outcomes. This developed system can also assist doctors in the field, supporting them to make decisions more easily and accurately.

## V. CONCLUSION

The model developed in this study has demonstrated high success, particularly in accurately identifying myocardial infarction and normal cases. However, challenges were encountered in classifying patients with a history of MI and abnormal heart rates, indicating the need for further improvements in the model's overall performance. Future work will focus on more extensive data processing and model optimization efforts to enhance accuracy and reliability for the less accurately identified classes. Such improvements will be crucial for the model's effective application in real-world clinical settings.